\documentstyle[12pt]{article}
\begin{document}

\begin{titlepage}
\begin{center}
\null\vskip-1truecm
\rightline{IC/94/168}
\vskip1truecm International Atomic Energy Agency\\
and\\
United Nations Educational Scientific and Cultural
Organization\\
\medskip INTERNATIONAL CENTRE FOR THEORETICAL PHYSICS\\
\vskip2.2truecm {\bf RELEVANCE OF INDUCED GAUGE INTERACTIONS\\
 IN
DECOHERENCE}
\vskip1.2truecm
Dhurjati Prasad Datta\footnote{\normalsize Permanent Address:
Department of Mathematics, North Eastern Regional Institute of
Science and Technology, Itanagar 791109, India.}\\  International Centre
for Theoretical Physics, Trieste, Italy. \end{center}
\vskip1.5truecm

\centerline{ABSTRACT}

\bigskip

Decoherence in quantum cosmology is shown to occur naturally in the
presence of induced geometric gauge interactions associated with particle
production. A new ``gauge''-variant form of the semiclassical Einstein
equations is also presented which makes the non-gravitating character of
the vacuum polarization energy explicit.

 \vfill

\begin{center}
 MIRAMARE -- TRIESTE\\
\medskip
July 1994
\end{center}
\vfill
\end{titlepage}
\newpage

Understanding the precise physical basis of a semiclassical regime of an
interacting quantum system is of great current interest [1].
A particular aim of the recent studies is to find a mechanism [2,3] which
should yield an almost classical observable from a given (nonclassical)
dynamical system. These have a special relevance also in quantum
cosmology (QC). Recently, there has been some attempts [4-11] in studying
the origin and validity of the semiclassical Einstein equations (SCEE) in
the framework of the quantum general relativity theory. The problem of
realizing a classical observable universe from a possible quantum regime
is also being addressed [5-7, 12] in the framework of the so called
``decoherence process'' in the quantum measurement theory [2,3].

In Ref. [8-10] the problem of obtaining a correct set of back reacted SCEE
is discussed in terms of the induced geometric gauge interactions.
Apart from yielding clarifications to some longstanding ambiguities [13]
in the semiclassical gravity, the study seems to offer new physical
characterizations of the gravitational effects of the quantized matter
fields.

It is well-known that geometric gauge interactions are induced
spontaneously on a heavy dynamical system when it is acted upon by a
light quantum system. In the language of the measurement theory, the
light system may be considered as an environment. The natural framework
of studying such an interaction is the Born--Oppenheimer (BO)
approximation modified by the induced gauge fields [14]. The influence of
the light system on the heavy one is succintly encoded in the form of the
induced gauge fields, besides the usual BO potential term [15], once the
light system is decoupled from the total system. The effects of these
induced gauge fields are known to yield interesting physically testable
predictions, e.g., the shift in   energy spectra of the heavy system [14].

In semiclassical cosmology, the effects of the induced gauge fields seem
to have deeper implications. One of the most important difference
between the ordinary quantum mechanics and the QC is that in QC the
universe is a closed system without a given a priori  time.
It is argued that the concept of time in cosmology may be an approximate
one, retrieved possibly  at a semiclassical regime. In such a case one
should be able to separate from among the total infinite number of degrees
of freedom, a smaller number of heavy degrees of freedom which behave
quasiclassically in the environment of the remaining lighter degrees of
freedom (matter + graviton). Although this separation seems to be
realizable  in the context of a minisuperspace, the physical implications of
the same turn out [10] to be nontrivial.

The definition of the semiclassical time is found to be intimately related
to the induced gauge structure, which in turn restricts severely the
character of the BO potential in the effective Hamiltonian of the heavy
degrees of freedom.  In fact, it is shown [10] that the BO potential in QC is
determined by the induced (vector) gauge connections so that the entire
influence of the environment is encoded in the induced gauge fields.
Physically, the environment leaves an imprint on the massive gravitational
degrees of freedom in the form of an induced gauge bundle, which is
nontrivial only if the gravitational minisuperspace has a nontrivial
geometric/topological structure. In a (Lorentzian) Robertson-Walker (RW)
minisuperspace e.g., the gauge bundle is trivial letting the associated
gauge connection and hence the BO potential (which is actually an energy
expectation value/transition matrix element) gauge equivalent to zero. An
implication of this result is that the vacuum energy in a RW universe must
be non-gravitating.
Although counter-intuitive, this seems to
 offer a natural resolution of the problem why the
cosmological constant in the present universe is negligibly small.

We note that the non-gravitating nature of
the vacuum energy is a consequence of the semiclassical definition of
 time as a parametric
derivative [6-11] which in turn relates it to the existence of a
geometric phase [8-10]. When the phase is zero (the RW case), time cannot
be defined at this level, reducing the (global) vacuum polarization energy
non-gravitating [10]. An interesting way of understanding this effect is
through the analogy of
the ordinary
electromagnetism. The global vacuum energy
carries an induced ``magnetic'' charge which does not contribute in the
first (gravitational) energy integral. An immediate problem is then to
suggest a new definition of time. This is however achieved by the next
lower order ``electric'' type back reaction of the particles created by the
time varying  gravitational background. The particle production effect is
associated with a nontrivial Pancharatnam phase [9]  which is pinned by
an Euclidean time integral of the total decay width of the unstable state.
In Ref. [10] we obtain the electric type potential from the particle
production effect and its back reaction in SCEE. The physical time is then
shown to have its origin from this Euclidean time by an analytic
continuation. Below we give further justification of this construction by
expressing the SCEE in a suitably defined new ``gauge''. However,
one still needs to clarify in more detail the exact meaning of the ``initial''
Euclidean time. A pertinent question is: what was, if any, before the
Euclidean era?

Further as stated already the origin of time needs the existence of a
nontrivial geometric phase which in turn needs to single out a
unicomponent complex WKB state in the gravitational sector [10,11]. One
should therefore also clarify the relevance of the Berry geometric phase
in the context of decoherence, which we do here.

The relevance of decoherence and correlations in QC in connection with
the back reacted SCEE is being studied by Hu and collaborators [4-7] for
sometime. Paz and Sinha [6] in particular have shown that the SCEE can be
derived by demanding decoherence in different WKB branches in the
universe wave function. The importance of achieving correlations (in the
sense of a gaussian peak in the Wigner function) between relevant physical
degrees of freedom in a single WKB branch is also emphasized in the
context of specific models. According to the decoherence paradigm [2] the
``relevant'' quantum degrees of freedom are constantly being measured
continuously by the environment degrees of freedom through some
nontrivial interaction. The environment degrees of freedom are however
not observable. These degrees of freedom should therefore be traced out in
the total density matrix of the system and environment. This leads to a
reduced density matrix of the system which includes the influence of the
environment on the system. The effects of environment on the system get
encoded in the Feynmann-Vernon (FV) influence functional [16] which
brings out the nonlocal character of the same. As noted in Ref. [6] the back
reactions turns out to depend on the histories of the system  variables.
The relevance of particle production in decoherence is also pointed out
[5,6]. Recently, the relation between this approach in QC and the quantum
Brownian problem has been stressed. Some interesting relationships
between dissipation; noise and the cosmological particle production have
been proposed [7].

However, this approach still seems to be incomplete since the possibility
of an induced gauge interaction in the Feynman-Vernon functional has not
been explored. Here, we show that decoherence is actually controlled by
the nontrivial geometric phase associated with particle production. A
 loss of quantum coherence in QC is thus related to the emergence of
a semiclassical time.
B
Consider a minisuperspace model with $N$ degrees of freedom denoted by
the coordinates
$q_i$ ($i = 1,\dots,N)$. The minisuperspace modes are coupled to the
environment degrees of freedom that are denoted here by a scalar field
$\varphi$. The quantum mechanical description of this interacting system
is governed by the Wheeler-Dewitt (WD) equation:
$$H\Psi \equiv (H_g+H_m)\Psi(q,\varphi)=0\eqno(1)$$
Here $\Psi(q,\varphi)$ is the total wavefunction of the universe filled
with matter $\varphi$. For definiteness, we consider the gravitational
Hamiltonian $Hg$ in the form
$$ Hg={1\over 2M} G^{ij} P_iP_j+MV(q)\eqno(2)$$
$G^{ij}$ denotes the metric in the minisuperspace, $P_i$ the momentum
conjugate to the configuration variables $q_i$, $V(q)$ the superpotential.
The quantity $M$ is proportional to the square of the Planck mass. The
Hamiltonian for the matter fields (environment) $H_m(\varphi,q)$ is kept
arbitrary. Because of the large value of $M$ compared to the ordinary
matter mass scales, one can use BO approximation to discuss the
solutions of (1).

To apply the improved BO approximation to Eq.(1) we write
$$\Psi =\psi(q)\chi(q,\varphi)\eqno(3)$$
and assume that the quasiclassical  gravitation modes $q$ are adiabatic
[17]. Then integrating the total Hamiltonian between an initial and a final
matter states $\chi_{i}$ and $\chi_f$ respectively, one gets the
effective gravitational WD equation
$$H_{eff}\psi(q) \equiv [{1\over 2M}G^{ij}(P_i-\hbar A_i)(P_j-\hbar
A_j)+MV+{\langle \chi_f\vert H_m\vert\chi_i\rangle\over
\langle \chi_f\vert\chi_i\rangle} +{\hbar\over 2M}
\rho]\psi = 0\eqno(4)$$
where the induced $U(1)$ gauge connection $A_i$ and the electric type
potential $\rho$ is given by [10,15]
$$A_i = i{\langle \chi_i\vert\partial_i\chi_i\rangle\over
\langle \chi_i\vert\chi_i\rangle},\quad
\partial_i={\partial\over\partial q_i}\eqno(5a)$$
$$\rho =\hbar G^{ij}{\langle \chi_f\vert(P_i-\hbar A_i)(P_j-\hbar
A_j)\vert\chi_i\rangle\over \langle \chi_f\vert\chi_i\rangle}\eqno(5b)$$
For details of the derivation of Eq. (4) we refer to Ref. [9].

Now a semiclassical regime is assumed to be one when the effective
gravitational wave function $\psi(q)$ can be approximated by an
oscillatory WKB state $\psi\sim \exp({i\over \hbar} S(q))$. In
this case the matter state $\chi(q,\varphi)$ corresponds to a curved space
Schr\"odinger wave functional
$$i\hbar {d\over dt} \vert\chi \rangle = H_m\vert\chi\rangle\eqno(6)$$
where the semiclassical WKB time is defined by
$${d\over dt}=G^{ij} {\partial S\over \partial q_i} {\partial\over\partial
q_j}\Rightarrow
{dqi\over dt}
=G^{ij} {\partial S\over \partial q_j} =G^{ij} {P_j\over M}\eqno(7)$$
At a particular value of the superspace variable $q$, $\vert\chi\rangle$
thus stands for a suitable Fock state constructed using the formalism of
the curved space quantum field theory. The choice of a vacuum, e.g., cannot
therefore be made unambiguously. However, under the assumed adiabatic
condition, one may be able to pick an adiabatic vacuum [18] which we do
here. For a general case special symmetry (e.g., de Sitter or conformal)
may need to be invoked. The choice of an adiabatic vacuum is however
sufficient for our purpose of showing the relevance of geometric phase in
decoherence.

The SCEE with back reactions now can be written as,
$${1\over 2M} G^{ij}(P_i-\hbar A_i)(P_j-\hbar A_j)+MV+{\langle
\chi_f\vert H_m\vert\chi_i\rangle\over
\langle \chi_f\vert\chi_i\rangle}+{\hbar\over 2M}\rho = 0\eqno(8)$$
where the classical effective momentum $P_i$ is defined by [c.f. Eq. (7)]
$$P_i=M{\partial S\over\partial q_i}\eqno(9)$$

We note that the transition matrix element
$\langle H_m\rangle =\langle \chi_f\vert H_m\vert\chi_i\rangle/\langle
\chi_f\vert \chi_i\rangle$ (expectation values in case
$\vert\chi_f\rangle
\equiv \vert\chi_i\rangle$) carries a geometric ``magnetic'' charge:
$$\langle H_m\rangle =\hbar A_i\ {dq_i\over dt}\eqno(10)$$
which follows from Eqs. (6) and (7). The last lower order term $\rho$ in Eq.
(8), however, corresponds to an ``electric'' potential [15] for a particle (in
minisuperspace) with an ``electric'' charge $\hbar/2M$. This also gives the
back reaction for the particles created on the background geometry. The
effect of renormalization on the vacuum transition element $\langle
H_m\rangle$ has been discussed in Ref. [19]. The local geometrical part of
the vacuum polarization contributes to the higher derivative corrections
in the Einstein-Hilbert action and has well defined gravitational effect.
(For simplicity we fix the renormalized higher derivative coupling
constants to zero). However, for a flat  simply connected minisuperspace
(e.g. the RW and most of the homogeneous models) the induced $U(1)$
bundle is trivial. Thus one can always make a gauge choice reducing the
global renormalized vacuum polarization component of $\langle
H_m\rangle$ to zero.

However, the adiabatic variation of the background geometry is expected
also
to produce nonperturbative effects in the matter sector. This induces
exponentially small imaginary part in the transition matrix element:
Im$\langle H_m\rangle =\Gamma$, the total decay width of the vacuum.
This lack of adiabaticity can however be treated in the adiabatic
perturbation theory using the Euclidean time formalism [9]. The Euclidean
integral
$$\gamma_P =\hbar^{-1}\int^f_i \Gamma d\tau,\quad \tau = it\eqno(11)$$
gives the nontrivial geometric Pancharatnam phase between the initial
and final vacua. In fact, the potential $\hbar \rho/2M$ is an effect of this
nonintegrable phase in the SCEE. The SCEE derived from a
simply-connected flat minisuperspace thus assumes the form [10]
$${1\over 2M} G^{ij}P_iP_j +MV +{\hbar\over 2M}\rho =0\eqno(12)$$
A remark on the definition of time is in order here.

The vacuum energy $E_0=Re\langle H_m\rangle$ being gauge equivalent to
zero, a physical time $t$ cannot be immediately written down via Eq. (7).
The WKB time should originally be retrieved as an Euclidean parameter
$${d\over d\tau} =G^{ij}{\partial S^E\over \partial
q_i}{\partial\over\partial q_j}\ ;\eqno(13)$$
where the Euclidean action $S^E=iS$. A simple way of recovering the
physical time $t$ is now to make a ``gauge'' transformation
$${d\over d\tilde \tau}=e^{\hbar^{-1}\int E_0 d\tau} {d\over d\tau}
e^{-h^{-1}\int E_0 d\tau}\eqno(14)$$
and then introduce the rotation $\tilde\tau = i\tilde t$. The Schrodinger
Eq. (6) then assumes the form
$$ i\hbar {d\over d\tilde t} \vert\chi\rangle = \tilde H_m\vert
\chi\rangle,\quad \tilde H_m = H_m-E_0\eqno(15)$$
The time derivative entering in momentum $P_i$ in Eq. (12) should also be
taken with respect to $\tilde t$ (this amounts to a gauge transformation
eluded already). Eqs. (13)-(15) and (12) now constitute the final form of
the SCEE and the matter Schrodinger equation. The virtue of this
``gauge''-transformed representation is that it makes manifest the
non-gravitating nature of the vacuum  energy. (We omit tilde henceforth).
The transformation (14) seems to eat up the apparent non-zero vacuum
polarization energy in Eq.(6).

We note that the solution $\psi(q)$ of the effective WD equation (4)
represents a class of $(N-1)$-parameter solutions $\psi_n(q)$ [6]. The
total wave function should therefore be written as a general superposition
of the form
$$\Psi(q,\varphi)=\Sigma \psi_n(q) \chi_n(q,\varphi)\eqno(16)$$
The definition of time (7) and hence the matter Schrodinger state
$\chi_n$ thus depends on the specific choice of the WKB branch
$\psi_n\sim \exp(iS_n/\hbar$). It is therefore desirable to justify the
choice of a single component WKB state $\psi_n$ for a semiclassical
description of the Universe. This will be achieved by showing that
decoherence occurs both between different WKB  branches and inside a
single WKB component. In fact, the appearance of a nontrivial geometric
phase in connection with particle production along each WKB branch seems
to destroy the quantum coherence of the  total quantum system.

As discussed by Kiefer [12] and Paz and Sinha [6] the problem of
decoherence in QC can also be addressed in terms of the FV
influence functional [16]. This is defined by introducing the reduced
density matrix of the universe
$$ \rho_r(q,q') =\int d\varphi \Psi^* (q,\varphi)\Psi(q',\varphi)\eqno(17)$$
when all the irrelevant environment degrees of freedom are integrated out.
Using Eq. (16), the reduced density matrix can be rewritten as:
$$\rho_r(q,q') =\sum_{n.n'} \psi^*_n(q) \psi_{n'}(q')F_{nn'}(q,q')\eqno(18)$$
where the FV influence functional is given by [12]
$$F_{nn'}(q,q') =\int d\varphi
\chi^*_n(q,\varphi)\chi_{n'}(q',\varphi)\eqno(19)$$
Note that the influence functional is in general a nonlocal functional of
histories $q(t)$ and $q'(t')$ of the semiclassical universes as predicted by
the WKB branches $\psi_n(q)$ and $\psi_{n'}(q')$ respectively. For the
same WKB branch $n=n'$, the different histories $q$ and $q'$ correspond to
interference between a possible expanding and collapsing mode. For
decoherence to occur both the off-diagonal terms $F_{nn'}(q,q)$ and
$F_{nn}(q,q')$ in the reduced density matrix should be exponentially
suppressed. The reduced density matrix $\rho_r$ will then be a sum of
non-interfering WKB branches indicating the onset of a classical era.

The close similarity of the integral (19) and the close-time-path
vacuum generating functional has been noted by Calzetta and Hu [20]
recently. The integral (19) in fact gives the transition of an in-vacuum,
prepared at a fictitious time $t=-\infty$, along the history $q$
upto $t=T$ and then return to the in-vacuum again at $t=-\infty$ along
$q'$. The matter field histories $\phi$ and $\phi'$ are supposed to match at
time $T:\phi(T)=\phi'(T)$. Let the common in vacuum be
$\vert\chi_0\rangle$. Then one has
$$ F_{nn'}(q,q') =\langle \chi_0\vert
\chi'_n\rangle_{q'}\langle\chi_n\vert\chi_0\rangle_q\eqno(20)$$
To choose an adiabatic in-vacuum we need to invoke a possible switching
off mechanism of the gravitational interaction at $t\to -\infty$ [6]. It is
well known [6] that this introduces restrictions on the class of
admissible super potentials $V(q)$. We note that relative to the in-vacuum
$\vert\chi_0\rangle$, the state $\vert\chi_n\rangle$ will in general be a
multiparticle state in the out region $T\to\infty$. The amplitude $\langle
\chi_0\vert\chi_n\rangle_q$ then gives a measure of the total shift in
the in-vacuum due to the adiabatic evolution along the history $q$. To
evaluate the amplitude we use the adiabatic theorem [14]:
$$  \ell n \langle \chi_0\vert\chi_n\rangle_q=\ell n \left({1\over
\vert 1 +i\beta_n\vert}\right)
-i\gamma_n-i{\hbar\over 2M} \int \rho_n dt\eqno(21)$$
where the lowest order phase correction $\gamma_n$ corresponds to the
geometric phase
$$\gamma_n = \hbar^{-1}\int E_n dt
+\hbar^{-1}\int\Gamma_nd\tau\eqno(22)$$

Here, the state $\vert\chi_n\rangle$ is assumed to be normalized
throughout the entire period of its evolution. Further, $E_n$ denotes the
instantaneous adiabatic energy, $\Gamma_n$ the decay  width of the state
[9] and $\rho_n$ an higher order correction to the geometric phase [14]. The
exponential decay of the state $\vert\chi_n\rangle$ is determined by
$\beta_n$ which is in general related to the decay width $\Gamma_n$ and
the energies associated to the states $\vert\chi_0\rangle$ and
$\vert\chi_n\rangle$.

The final form of the influence functional can now be written as
$$ F_{nn'}(q,q')=\exp\left[i\left\{\hbar^{-1}(\int E_n dt -\int
E_{n'}dt')+\hbar^{-1}(\int\Gamma_n dt-\int\Gamma_{n'}dt')\right.\right.$$
$$
\left.\left.+{\hbar\over 2M} (\int \rho_n dt-\int\rho_{n'}
dt')\right\}\right]\times\exp[-(\beta_n-\beta_{n'})^2] \eqno(23)$$
In view of the fact that $\beta_n\propto\Gamma_n$ in the adiabatic
approximation, one concludes that the expected decoherence is obtained
whenever there is a large difference in geometric phases (associated with
particle production) along the two histories.
Further, the geometric phases associated with distinct histories ought to
be different. Otherwise the histories will be homotopically equivalent,
which follows from the topological invariance of the geometric phase.
 This
phase difference indicates varying rates of particle production along
different WKB histories, suppressing the off-diagonal terms. Stated in
other words, the loss of quantum coherence in WKB branches can be seen as
due to the lack of coherence in the induced geometric phases.
 This seems to be a new
interpretation of the relationship between decoherence and cosmological
particle production noted already by many authors [5-7].

In the case of a RW universe, the energy integrals in Eq.(23) can be gauged
away [WKB time $t(t')$ in this case is identical to one in Eq. (15)]. The
simplified influence functional then has the form
$$
F_{nn'}(q,q')=\exp[i\{\hbar^{-1}(\int\Gamma_nd\tau-\int\Gamma_{n'}d\tau')+{
\hbar\over
2M} (\int \rho_n dt -\int \rho_{n'}dt')]]$$
$$\times\exp(-(\beta_n-\beta_{n'})^2)\eqno(24)$$
One notes the interesting functional relationship between the phase and
the damping factor of the influence functional. The explicit derivation of
the cosmological fluctuation-dissipation theorem [7] in the present
context will be taken up separately.

Finally, as an application let us compute the influence functional damping
factor in a toy model. The model consists of a RW minisuperspace coupled
to a conformal matter field [5,6]. Since detailed calculations are available
in literature we will be brief bringing out the main features of the
present approach.

The relevant WD equation has the form
$$[{1\over 2M}{\partial^2\over\partial a^2}+MV(a)+{1\over
2}\sum_k(-{\partial^2\over\partial\varphi^2_k}+\Omega^2_k(a)
\varphi^2_n)]\psi=0\eqno(25)$$
Here, $\varphi_k$ are the scalar field eigenmodes of the spatial Laplacian
$\Delta$ on the unit 3-sphere, $\Omega^2_k=k^2+m^2a^2$, $m$ the  mass
of the scalar field and $a$ the RW scale factor. The matter wave function
$\chi(a,\varphi)$ turns out to be seperable
$$\chi(a,\varphi)=\mathop{\Pi}\limits_k\chi_k(a,\varphi_k)\eqno(26)$$
where each component $\chi_k$ satisfies the Schrodinger equation
$$i\hbar{d\over dt} \chi_k=(-{1\over
2}{\partial^2\over\partial\phi^2_k}+\Omega^2_k
\varphi^2_k)\chi_k\eqno(27)$$
and $t$ is the WKB time defined by ${da\over dt} ={dS\over da}$. Let
us assume a Gaussian ansatz for the wave function $\chi$ (we omit the
index $k$ for simplicity)
$$\chi(a,\varphi)=({\varepsilon\over\pi})^{1/4} e^{i\alpha(a)-{1\over
2}\sigma(a)\varphi^2}
\eqno(28)$$
where
v$$\sigma =\varepsilon +i\Gamma$$
$$2\hbar\dot\alpha =-\varepsilon\ {\rm and}\ \hbar\dot\sigma
=-i\sigma^2+i{\Omega^2_k}\eqno(29)$$
so that $\langle \chi\vert\chi\rangle = 1$.
The initial state $\vert\chi_0\rangle$ thus turns out to be the harmonic
oscillator ground state with energy $\varepsilon_0 ={1\over
2}\Omega(0)\hbar$. It is now straightforward to calculate
$$\langle \chi_0\vert\chi\rangle = e^{-i\Gamma/
4\varepsilon_0}
\bigg\vert1+i{\Gamma\over 2\varepsilon_0}\bigg\vert^{-1/2}\eqno(30)$$
where we use first order adiabatic correction in evaluating the wave
function $\vert\chi\rangle$ [6]. We also absorb the phase contribution
from the energy integral in the definition of time (c.f. Eq. (15)). Note,
however, the geometric phase $\Gamma/4\varepsilon_0$ associated to the
decay of the state. The influence functional then assumes the form
(inserting the index
$k$)
v$$F(a,a')
=\exp[i\Sigma{1\over
4\varepsilon_{0k}}(\Gamma_k-\Gamma'_k)]\exp[-\Sigma{1\over
8\varepsilon^2_{0k}}(\Gamma_k-\Gamma'_k)^2]\eqno(31)$$ This should be
compared with the expression obtained by Paz and Sinha [6]. Although the
damping factor is identical, the influence phase is different. One reason
for the difference is the non-gravitating nature of the vacuum polarization
energy.

To conclude, we have shown the close proximity of decoherence with the
occurrence of the nontrivial geometric phase in an interacting system. In
the cosmological context discussed here, decoherence of the WKB branches
of the universe histories is realized provided there is a large uncertainty
in the induced phase differences, reflecting a vastly different rates in
particle production. However, our result is still not sufficiently general
because of the adiabatic switching off mechanism used here. It is
desirable to remove this restriction from the argument. Finally, we have
also presented a new ``induced gauge'' variant form of the SCEE, rendering
the non-gravitating character of the global renormalized vacuum energy
manifest. These results seem to have interesting implications in quantum
gravity and quantum measurement problems.
\vspace{1.5cm}

\noindent{\large\bf Acknowledgments}
\bigskip

It is a pleasure to thank Professor S. Randjbar-Daemi, Professor A.O.
Caldeira and Professor Deepak Kumar for fruitful discussions, Dr. L.
Palacios for numerous help and constant encouragement, and Dr. C. Kiefer
for a
number of stimulating correspondence over electronic mail.  The author
would also like to thank Professor Abdus Salam, the International Atomic
Energy Agency and UNESCO for hospitality at the International Centre for
Theoretical Physics, Trieste.
v
\end{document}